# Gaussian Content as a Laser Beam Quality Parameter

Shlomo Ruschin,[1,2] Elad Yaakobi[2] and Eyal Shekel[2]

[1] *Department of Physical Electronics, School of Electrical Engineering Faculty of Engineering, Tel-Aviv University, Tel-Aviv 69978 Israel*

[2] *Civan Advanced Technologies, 64 Kanfei Nesharim street Jerusalem 95464, Israel*

[*]*Corresponding author: ruschin@eng.tau.ac.il*

*We propose the Gaussian Content as an optional quality parameter for the characterization of laser beams. It is defined as the overlap integral of a given field with an optimally defined Gaussian. The definition is specially suited for applications where coherence properties are targeted. Mathematical definitions and basic calculation procedures are given along with results for basic beam profiles. The coherent combination of an array of laser beams and the optimal coupling between a diode laser and a single-mode fiber (SMF) are elaborated as application examples. The measurement of the Gaussian Content and its conservation upon propagation are experimentally confirmed.*

# 1. Introduction

The issue of characterizing the beam quality of a coherent beam has been a matter of study and discussion for more than two decades [1,2]. Although several ways to characterize and evaluate the quality of coherent beams have been proposed, there is a widespread consensus that there is no single accepted beam quality parameter suitable for all applications and scenarios, and therefore, specific quality factors are often chosen by practical considerations and even "target oriented". Examples of established parameters [1] are the $M^2$ factor (or BPP-Beam Propagation Product related to it), the PIB ("Power in the bucket") and Strehl ratio parameters. The first one gives a good measure on the localization of the power while propagating, and has the advantages that it allows a simple description of the main power transport and facilitates basic design by means of the ABCD matrices formalism [3]. Presently the $M^2$ factor gained recognition as the best acknowledged quality parameter for laser beams and even acquired industrial standardization [4]. In some instances however the $M^2$ parameter happened to be inadequate for describing the suitability of a beam for important applications, the perhaps best known example being a flattened or "top-hat" beams. Such beams, although desirable in many applications where uniform coherent illumination is required, feature high $M^2$ values due to their steep power descent at the edges. An additional example is the combination of different coherent beams into one single source [5-7], a subject attracting much activity nowadays in order to enlarge source power retaining coherence properties. The unsuitability of the $M^2$ parameter in this context was pointed out recently by Zhou et Al. [7]. They proposed to rely on a different quality measure, based on PIB.

In this article an additional alternative definition of a beam quality factor is proposed. It defines in some sense the "best closeness" of a given beam to a standard Gaussian distribution, and it is suited to applications where the coherence properties of light at the target of preference. Examples of such applications are the coherent combination of different laser sources and the optimal coupling from a laser to a single-mode optical fiber.  In that sense, the new definition can be considered like other as target oriented, but in addition to it, the proposed parameter is a basic property of the entire beam and is preserved on propagation through homogeneous media and ideal lenses (first-order optical systems). As such it should be helpful in the description and design of optical systems by simple *ABCD* matrix formalism.

The continuation of this article is organized as follows: In the next section the *GC* parameter is defined and some mathematical properties and physical consequences following the definition are discussed. Section 3 contains several examples of simple profiles for which the *GC* is calculated and its value compared to the corresponding $M^2$ value. Two more elaborated examples follow, namely an array of coherent sources considered as a single beam and the optimal coupling between a diode laser and an SMF. Experimental results are brought in Section 6 where we demonstrate the extraction of the *GC* from measured data and its conservation upon propagation.

## 2. Mathematical Formulation

The mathematical basis of the present definition originates from the expansion of an arbitrary normalized coherent field *U(x)* by a set of complete orthonormal functions, which for definiteness will be chosen as the Gauss-Hermite set *{GH$^w_{n,m}$}*:

$$U(x,y) = \sum_{n,m} A^w_{n,m} GH^w_{n,m}(x,y) \quad , \quad A^w_{n,m} = [U(x,y) \cdot (GH^w_{n,m}(x,y))^*] \tag{1}$$

where the dot appearing in the expansion constant definition means scalar product given by the corresponding overlap integral, and $w$ is the Gaussian parameter that defines the set. One notices first that there are an infinite number of such expansions, as the parameter $w$ can take any real positive value. We look now at the lowest-order expansion coefficient $A_{0,0}^w$: The absolute value squared of this factor is upper-bounded as a function of $w$, ($|A_{0,0}^w|^2 \leq 1$), and as such it must reach an absolute maximum within the range $0 < w < \infty$. We call that maximum value the *Gaussian Content* (GC) of the beam, and the Gaussian parameter value for which this maximum is acquired, we designate by $w_{opt}$. We generalize the definition to allow for a spherical phase and define explicitly the *GC* factor as:

$$GC \equiv Max_{w,R} \left| \left[\frac{2}{\pi}\right]^{1/4} \frac{1}{w^{1/2}} \int_{-\infty}^{\infty} [U(x)]^* \exp\left(-i\frac{kx^2}{2}\left(\frac{1}{R} - i\frac{2}{kw^2}\right)\right) dx \right|^2 \qquad (2)$$

From this definition an optimal curvature radius ($R_{opt}$) also follows. The definition above applies to 1D beams and its extension to 2D is obvious. The following properties are a direct consequence of the *GC* being one of the expansion coefficients of the beam as defined in Eq. (1):

1. The parameter GC is positive and bound ($0 \leq GC \leq 1$) for all normalizable beams, and *GC=1* will occur only for a pure Spherical-Gaussian beam.
2. The value of GC will not change as the beam propagates through a first-order optical system, e.g. homogeneous space or elements that can be described by a real ABCD matrix.
3. The optimal Gaussian defined by ($w_{opt}, R_{opt}$) will transform according to the ABCD Gaussian law and optimize the transformed beam in the sense described above at any plane in the propagation path, including the target plane.

Item 2 above will hold for any optical element or beam transformation which preserves the absolute value of the expansion coefficients defined in Eq. 1. The GC can be given also a tangible physical interpretation: Keeping in mind that the overlap integral evaluates also the coupling efficiency between a waveguide mode and an input beam, and in view of consequences 1-3 above, one can define the GC in physical terms as follows: *The GC parameter is the optimal coupling efficiency that can be achieved between a given field and a Gaussian mode of a waveguide or optical fiber by means of first-order optical elements that can be expressed by real ABCD matrices.* Standard SM optical fibers support a mode which is very well approximated by means of a Gaussian [8], and can be used for a practical test.

## 3. Calculation examples

In the examples following we simplify somewhat the calculations by assuming a beam with a planar phase ($R = \infty$). The field function $U(x)$ is then real, and the maximalization expressed in Eq. (2) can be performed simply by its differentiation with respect to $w$. The following "working equations" are then attained for the determination of $w_{opt}$ and GC:

$$\int_{-\infty}^{\infty} U(x) \cdot \exp[-(x/w_{opt})^2 (2(x/w_{opt})^2 - 1/2]dx = 0 \tag{3}$$

$$GC \equiv \left[\frac{2}{\pi}\right]^{1/2} \frac{1}{w_{opt}} \left|\int_{-\infty}^{\infty} U(x) \exp\left(-\left(\frac{x^2}{w_{opt}^2}\right)\right) dx\right|^2 \tag{4}$$

Equation (3) is an algebraic equation with a single unknown ($w_{opt}$), and Eq. (4) is evaluated once the optimal Gaussian parameter $w_{opt}$ is determined. It is worthwhile to remark that these two last equations have both integral formulations and as such, they are solvable in principle for any square-integrable test function $U(x)$, making the determination of the quality factor GC immune to possible discontinuities in *U(x)* or its derivatives. The planar-phase assumption is removed in the experimental results analysis of Section 5. For some simple beam profiles *U(x)*, the integrals in Eqs. (3) and (4) can be explicitly solved.

Turning to specific examples, a straightforward implementation of Eqs (3) and (4) will render for a rectangular top-hat function of half-width *a*, a GC value of 0.89 and an optimal Gaussian width $w_{opt}$ = 1.01*a*. A similar calculation for a 2D radial top-hat of radius *a*, will furnish values of *GC* = 0.815 and $w_{opt}$ = 0.89*a* . Additional values of *GC* and $w_{opt}$ for simple profiles are given in Table I. Corresponding $M^2$ values are also shown for comparison. The half-cosine shape listed in the table is an approximation of the field exiting a basic mode in a high-contrast waveguide, and the exponential one has been proposed as a model for a thin diode-laser output.

Table I: Gaussian Content, $w_{opt}$ and $M^2$ values of simple beam profiles

| Optical Field | Shape | $W_{opt}$ | GC | $M^2$ |
|---|---|---|---|---|
| Gaussian<br>$U(x)=exp[-(x/w_G)^2]$ | 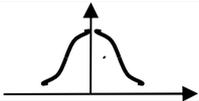 | $w_G$ | 1 | 1 |
| Rect(1D)<br>$U(x)=\Phi(x-a)-\Phi(x+a)$ | 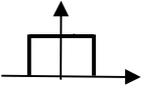 | 1.01*a* | 0.89 | ∞ |
| Top-hat(2D)<br>$U(r)=\Phi(r/a)$ | 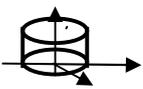 | 0.892*a* | 0.815 | ∞ |
| Exponential<br>$U(x)=exp(-|x|/\varepsilon)$ | 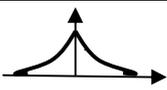 | 1.31*ε* | 0.972 | 1.414 |
| Half-cosine<br>$U(x)=cos(\pi x/2a)$ | 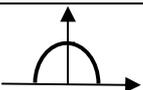 | 0.703*a* | 0.99 | 1.136 |

In the next example, we calculate *GC* and $w_{opt}$ for Super-Gaussian beams which have been widely implemented as a model for gradual transition between an ideal Gaussian and a top-hat beam. Explicit calculations of the $M^2$ parameter for this type of beams were given in ref. [9]. The comparison between calculated *GC* and $M^2$ for Super-Gaussian functions of order *n*, in the 1D case, is plotted in Fig.1. It is seen how the $M^2$ parameter diverges while GC remains basically unchanged from *n ~ 20* on.

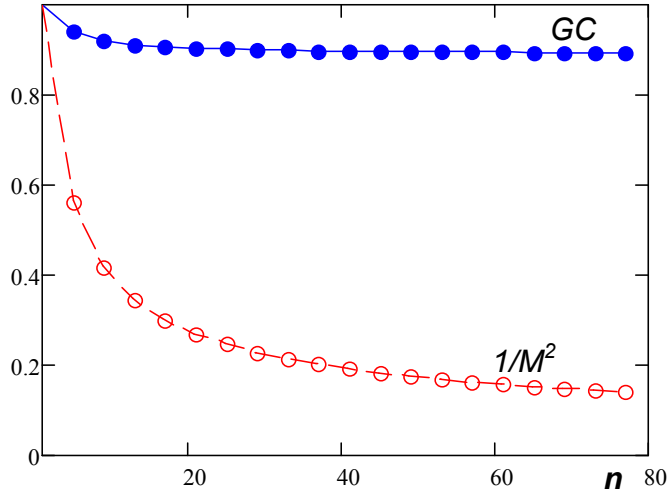

Figure 1. *GC* and *$1/M^2$* parameters for Super-Gaussian beams as a function of their order *n*. The $M^2$ parameter was inverted in order to fit the same graph.

If the tested distribution is displaced off-axis with respect to the fitting Gaussian, the GC parameter will be reduced. Indeed *GC* and $w_{opt}$ can be analytically calculated for an off-axis Gaussian displaced by an amount *d* to give:

$$GC = \frac{2w_{opt}wg}{w_{opt}^2 + wg^2}\exp(\frac{-2d^2}{w_{opt}^2 + wg^2}) \qquad (5)$$

$$w_{opt} = [2d^2 + (4d^4 + wg^4)^{1/2}]^{1/2}$$

The exact position of the optimal Gaussian center with respect to the tested beam has to be taken therefore into account for general beams, and determined either by symmetry, geometrical considerations or as an additional parameter for optimizing the *GC*.

### 3.1 Study case: A periodic linear array of Gaussian sources

As a further example of the applicability of the Gaussian Content factor in a practical system, we characterize the GC of an optical source composed by a periodic array of N beams disposed on a line. Coherent combination of laser beams, as mentioned, is a timely topic, and the need to define a proper parameter for its quality evaluation has been recently addressed [10]. A typical input field is seen in Figure 2. It is basically characterized by 3 parameters, namely the number of beams $N$, the width of a single Gaussian $w_g$ and the period $\Lambda$. In the same graph we also draw the Best-fit Gaussian (BFG) of the array.

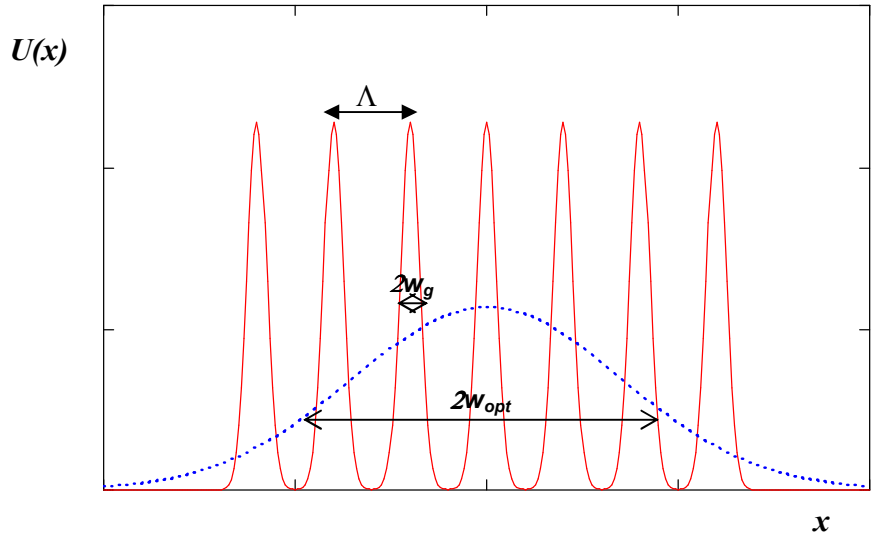

Fig.2. Input field consisting of 7 beams of width $2w_g$ and period $\Lambda$. The dotted line shows the best-fit Gaussian (BFG)

The next graph shows the GC as a function of the number of beams in the array, for different values of the inverse filling factor $\Lambda/w_g$. As seen the GC rapidly converges to a constant value, and becomes independent of the number of beams in the array. This property is also strikingly distinct from the $M^2$ parameter which increases quasi-linearly with the number of element in the array [11]. Choosing a specific case: a linear array of 9 Gaussians with $\Lambda/w_g = 2$ will feature a $M^2$ of about 5, which would correspond to a low quality coherent source. The *GC* for that configuration on the other hand, as depicted in Fig. 3, turns out to be 0.895 meaning that such an array is able to couple almost 90% of its power into a single-mode fiber using first-order optical elements. Another point to remark in the graph of Figure 3, is that for high values of the ratio (e.g. $\Lambda/w_g = 20$ in Fig. 3), two local maxima are found for $w_{opt}$, denoted by the dotted lines in the graph. As seen, the absolute maximum is chosen, and the transition between the two solutions takes place at $N = 9$.

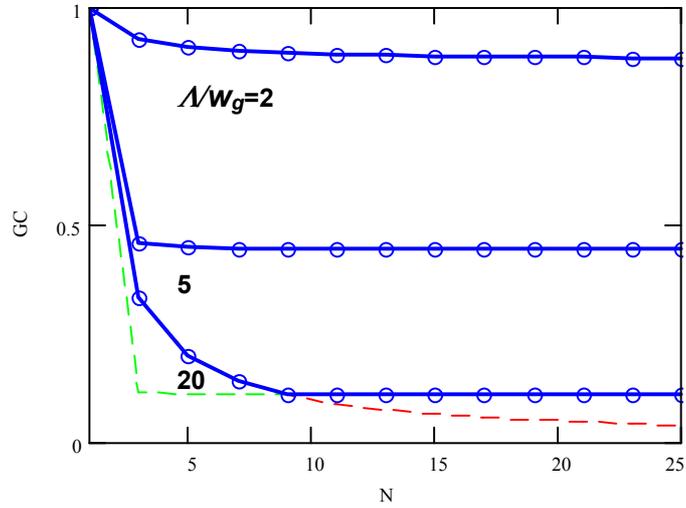

Figure 3. Gaussian content as a function of beams in the array, for different values of the normalized period $\Lambda/w_g$. The dotted lines for $N=20$ (green and red on-line) denote local maxima of the overlap integral from which the *GC* is chosen.

**3.2 Study case: Optimal coupling between a diode laser and single-mode fiber**

As mentioned in the previous section the GC parameter is suited by its definition to be applied to the problem of optimal coupling between a single-mode laser and a SMF. This optimization problem has been intensively studied and a few references are quoted [12-16]. No systematic design of such a coupling system will be attempted here, but merely we mention some basic conclusions on the attainable efficiency following the *GC* definitions and properties presented so far.

A simple proposed way to model the field distribution at the exit facet of a thin diode laser was proposed in terms of the Exponential-Gaussian 2D profile [12-14]:

$$U(x,y) = \exp\left(-|x|/\varepsilon\right)\exp(-\frac{y^2}{2\sigma^2}+\frac{iyx^2}{2R}) \quad (6)$$

This profile was analytically studied and optimized by Serna et. Al.[14], using a model based on second order-moments. Since the profile of Eq. (6) is expressed as factorized functions in the dimensions *x* and *y*, both, the 2-dimensional M2 and *GC* parameters are given by the product of the corresponding 1D factors. Since the profile is assumed Gaussian in the *y* direction, obviously: $GC_y = M_y^2 = 1$. On the other hand a straightforward calculation based on Eqs.(3) and (4) for the exponential profile, renders $w_{opt}$ = *1.31ε* and a Gaussian Content $GC_x$ = *0.972* . This number sets an upper limit for the efficiency of power coupling of all DL sources modeled by (5) of 97.2%. Furthermore, this value can be ultimately approached for such by a set of ideal orthogonal-cylindrical lenses designed for properly evolving the optimized Gaussians. Indeed, *GC* values exceeding 0.93 were measured by us as described in the experimental section below. It is instructive to compare these results with the analysis in [14]: The authors there defined a Beam Quality parameter *Q* which is actually a generalization of the $M^2$ parameter for the case of a non-circularly symmetric beam. Optimization of the quality parameter defined there (in terms of σ and ε) was attained demanding second-moment equalization in the two orthogonal directions and furnished: $\varepsilon/\sigma = 2^{1/4} = 1.19$. For comparison, based on *GC* calculation, without implementation of astigmatic optics, the best 2D coupling efficiency (0.92) will be obviously met if $2\sigma^2 = w_{opt}^2 = 1.31^2\varepsilon^2$, meaning an optimal ratio of $\varepsilon/\sigma = 1.08$. The analysis based on the moment method furnished an upper limit for the DL quality of $Q = 1.46/k^2$. A value of $M_y^2 = 1.41$ can be also inferred for profile (5) by direct implementation of the formalism given in [9].

## 5. Experimental results

The equipment needed in order to determine the *GC* is basically a good quality detector array, camera or a scanning device. In order to confirm the theoretical predictions and the suitability of the GC as FOM for characterizing laser beams, we implemented the method on two semiconductor sources of different quality fabricated by the same process and emitting at a wavelength of λ = 975nm. The fast axis of the source was collimated by a first cylindrical lens, while a second cylindrical lens placed orthogonally to the first one was used in order to focus the light in the (slow) horizontal direction. The beam were characterized by means of a commercial beam profiler (Spiricon model SP620U), and scans were taken in the horizontal direction near-field in the vicinity of the focal plane. The calculation of the *GC* parameter requires the knowledge of the complex field distribution and the required phase retrieval was accomplished by means of a standard Saxton-Gerchberg algorithm [17] based upon three consecutive intensity scans. The GC was calculated for two sources, one of good quality and one for low quality, yielding values of 0.93 and 0.52 respectively. In order to confirm the conservation of the GC factor on propagation, the GC was further calculated *independently* for more than 100 consecutive profiles along the propagation path by means of the optimization procedure described above, concurrently with a calculation based on standard Fresnel diffraction integral. The results are shown in Fig. 4 and 5. As seen, the values of the GC remain basically consistent for all scanned profiles. This consistency is especially remarkable for the low-quality beam, which changes its profile in an abrupt way while propagating. Even in that situation our method was still able to extract the common hidden optimal Gaussian and corresponding GC parameter confirming the robustness of the GC calculation procedures. In Fig 4 and 5 we see also the propagation of the optimal Gaussian width $w_{opt}(z)$ and spherical phase radii $R_{opt}(z)$ as a function z as calculated in each plane by GC optimization. The propagation values are compared to the usual Gaussian formula and the fit is very good.

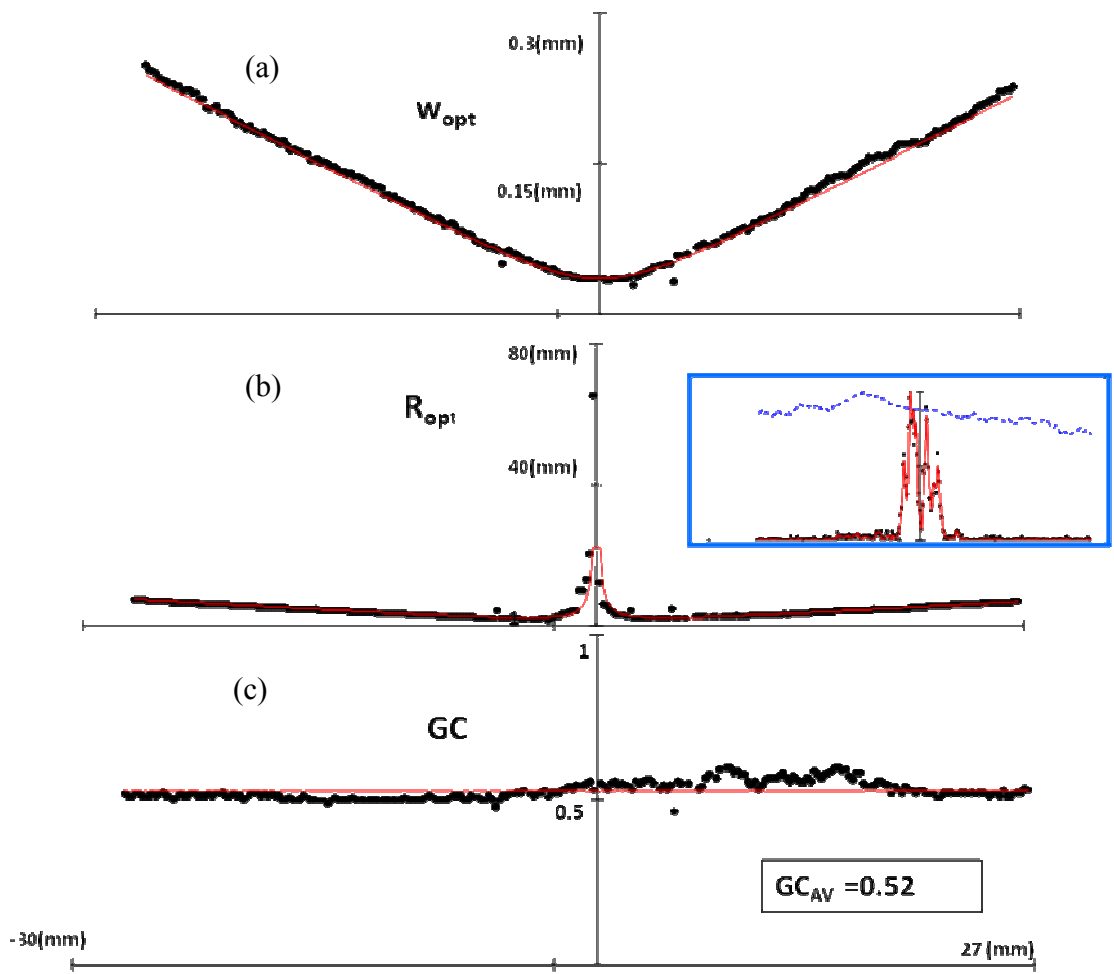

Figure 4. Parameters of the optimized Gaussian and Gaussian Content computed independently at more than 100 planes in the vicinity of the focal plane as a function of z for a low quality beam. (a) The Gaussian width w(z), (b) The spherical phase radius of curvature R(z), (c) The Gaussian Content (*GC*) parameter. The red lines correspond to standard Gaussian fit. The inset shows a sampled profile (intensity and phase) near the waist of the beam.

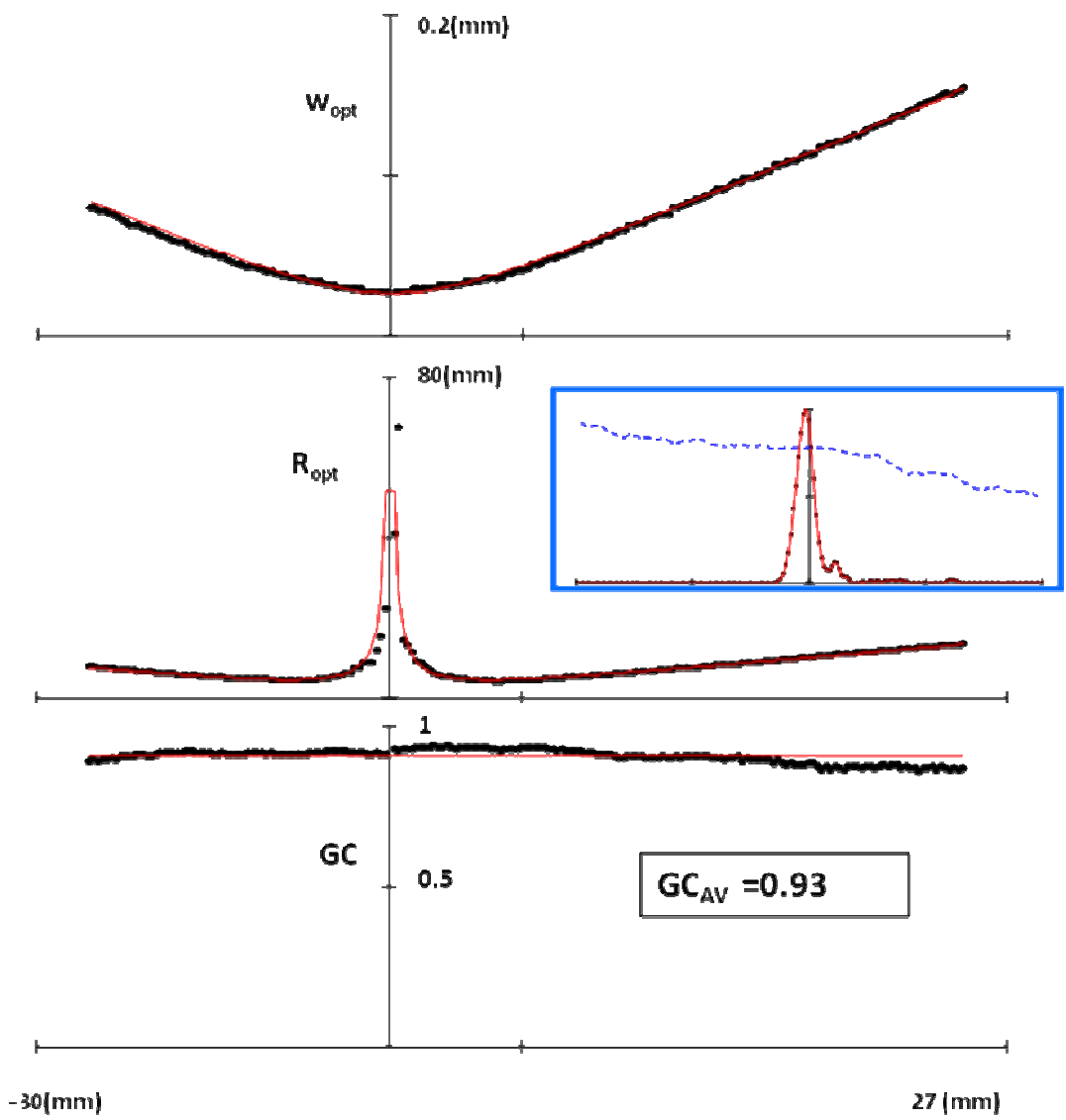

Figure 5. Similar to Figure 4, but measuring a higher quality beam.

## 4. Conclusions

An alternative beam quality parameter was defined and demonstrated. As with other definitions, its choice among other options will be determined by the targeted application. Specifically we propose the *GC* as a figure of merit of preference for applications where coherence properties at the target are of priority. As application examples for the *GC* FOM, we suggested the coupling of a laser source into a single-mode fiber and the coherent combination of a beam array source. In this presentation, we have limited the analysis to coherent fields that can be described by the paraxial scalar approximation. These assumptions apply to many customary sources and could be eventually removed by further analysis. The GC determination was experimentally demonstrated for a semi-conductor laser source and its conservation upon propagation verified for both low quality and high quality beams. The experimental procedure required for the evaluation of the GC parameter was based on the same hardware and similar raw data as that required for characterization by moment methods.

In addition to its potential usefulness as coherent source FOM, the GC method can be applied as a design aid in optical systems where coherence is a priority target, e.g. interferometric setups and coherent beams combiners. As defined, the *GC* method extracts from the beam a specific part of it in terms of an ideal coherent (Gaussian) beam. This *Best Fit Gaussian* is characterized by its width $w_{opt}$ and curvature radius $R_{opt}$. For important coherent applications this underlying Gaussian represents the "useful" part of the incoming beam and the optical design may be aimed at its optimal delivery disregarding the rest of the beam. First-order optical elements preserve the value of *GC* and the optimal Gaussian parameters ($w_{opt}$, $R_{opt}$) transform along propagation according to simple *ABCD* laws. Other type of elements, e.g. non-spherical, lenslet arrays or diffractive-optical, can change and even improve the value of the *GC* parameter. In that case, the GC can act as target parameter in order to evaluate those schemes.

**References**

1. H. Weber "Some historical and technical aspects of beam quality" Optical and Quantum Electronics **24,** 861-864 (1992)

2. A. E. Siegman, "Defining, measuring, and optimizing laser beam quality", Proc. SPIE **1868**, 2 (1993)

3. N. Hodgson and H. Weber, *Laser resonators and beam propagation: fundamentals, advanced concepts and applications*,(Springer, 2005*)*

4. ISO Standard 11146, "Lasers and laser-related equipment – Test methods for laser beam widths, divergence angles and beam propagation ratios" (2005)

5. T.Y Fan, "Laser beam combining for high-power, high-radiance sources" , IEEE J. Sel. Top. Quantum Electron. **11**, 567 (2005).

6. R. Xiao, J. Hou, M. Liu, and Z. F. Jiang " Coherent combining technology of master oscillator power amplifier fiber arrays", Optics Express, 16, pp. 2015-2022 (2008)

7. P. Zhou, Z. Liu, X. Xu, Z. Chen and X. Wang "Beam quality factor for coherently combined fiber laser beams", Optics &Laser Technology **41**(2009)268–271

8. D. Marcuse, "Gaussian approximation of the fundamental modes of graded-index fibers" J. Opt. Soc. Am., **68**, Issue 1, pp. 103-109 (1978)

9. P. A. Belanger and C. Paré, "Optical resonators using graded-phase mirrors", Optics Lett. **14,** 1057-1059 (1991)

10. P. Zhou, Y. Ma, X. Wang, H. Ma, J. Wang, X. Xu, and Z. Liu, "Coherent beam combination of a hexagonal distributed high power fiber amplifier array", Appl. Optics **48** 6537-6540 (2009)

11. X. Ji, T. Zhang and X. Jia " Beam propagation factor of partially coherent Hermite–Gaussian array beams" J. Opt. A: Pure Appl. Opt. 11, 105705 (2009)

12. D.D. Cook and F. R. Nash, J. Appl. Phys. 46 (1975) 1660.

13. W.P. Dumke, "The Angular Beam Divergence in Double-Heterojunction Lasers with Very Thin Active Regions" IEEE J. Quantum Electron. QE-11 (1975) 400-402.

14. J. Serna, P. M. Mejías and R. Martínez-Herrero "Beam quality in monomode diode lasers " Optical and Quantum Electronics **24** 881-887 (1992)

15. G. Hunziker and C. Harder "Beam quality of InGaA sridge lasers at high output power" Appl. Optics **34** 6118-6122 (1995)

16. W. D. Herzog, M. S. Unlu ,B. B. Goldberg, G. H. Rhodes and C. Harder "Beam divergence and waist measurements of laser diodes by near-field scanning optical microscopy" Appl. Phys. Lett. **70** 688-690 (1997).

17. J. R. Fienup, "Phase retrieval algorithms: a comparison" Appl. Optics **21** 2758-2769 (1982)